\documentclass{xjkas}

\def\year{2015} 
\def\month{June} 
\def\volume{00} 
\def\beginpage{1} 
\def\endpage{4} 
\def\received{May 28, 2015} 
\def\accepted{June 15, 2015} 
\date{Received \received; accepted \accepted}

\usepackage{flushend} 
\newcommand\ion[2]{{#1}\,{\sc #2}} 
\usepackage[breaklinks,colorlinks,citecolor=blue,linkcolor=blue,menucolor=blue,urlcolor=blue]{hyperref}

\def\Ll{L_{\lambda}}
\def\lLl{\lambda L_{\lambda}}
\def\LC{L_{1350}}
\def\LH{L_{5100}}
\def\Haa{_{\rm H\alpha}}
\def\Hbb{_{\rm H\beta}}
\def\H{_{\rm H}}
\def\C{_{\rm C\,IV}}
\def\Ha{H$\alpha$}
\def\Hb{H$\beta$}
\def\Msol{M_{\odot}}

\title{
AGN Broad Line Regions Scale with Bolometric Luminosity\thanks{\sc Rapid Communication}
}

\author[]{Sascha~Trippe}

\affil[]{Department of Physics and Astronomy, Seoul National University, Seoul 151-742, Korea; \email{trippe@astro.snu.ac.kr}}

\def\corrauthor{
S. Trippe
}

\def\runningauthor{
Trippe
}

\def\runningtitle{
AGN Broad Line Regions Scale with Bolometric Luminosity
}

\def\keywords{
galaxies: active --- quasars: emission lines --- black hole physics
}

\def\abstracttext{
The masses of supermassive black holes in active galactic nuclei (AGN) can be derived spectroscopically via virial mass estimators based on selected broad optical/ultraviolet emission lines. These estimates commonly use the line width as a proxy for the gas speed and the monochromatic continuum luminosity, $\lLl$, as a proxy for the radius of the broad line region. However, if the size of the broad line region scales with the \emph{bolometric} AGN luminosity rather than $\lLl$, mass estimates based on different emission lines will show a systematic discrepancy which is a function of the color of the AGN continuum. This has actually been observed in mass estimates based on \Ha/\Hb\ and \ion{C}{iv} lines, indicating that AGN broad line regions indeed scale with bolometric luminosity. Given that this effect seems to have been overlooked as yet, currently used single-epoch mass estimates are likely to be biased.
}

\begin{document}
\jkashead 


\section{Introduction \label{sec:intro}}

Supermassive black holes, with masses ranging from millions to billions of solar masses, are components of probably all galaxies \citep[e.g.,][]{fletcher2003, ferrarese2005}. Accretion of gas onto such black holes gives rise to the phenomenon of active galactic nuclei (AGN) which in turn are able to influence the evolution of their host galaxies via ``AGN feedback'' \citep[e.g.,][]{fabian2012}. Accordingly, knowledge of the masses of supermassive black holes is of great astrophysical importance.

For nearby galaxies, black hole masses $M$ can be derived from the dynamics of stars \citep[e.g.,][]{bender2005, oh2009}, of gas \citep[e.g.,][]{walsh2013}, or of interstellar masers \citep[e.g.,][]{herrnstein2005}. For distant AGN, black hole masses can be derived from spectroscopy of broad optical/ultraviolet emission lines. Such mass estimates make use of a virial relation of the form
\begin{equation}
M = f\,\frac{v^2 R}{G}
\label{eq:virial}
\end{equation}
where $v$ is the root-mean-squared line-of-sight gas velocity, $R$ is the (effective) radius of the broad line region, $G$ is Newton's constant, and $f$ is a geometry factor of order unity. The velocity $v$ is derived from the widths of the emission lines. The radius $R$ can be derived via reverberation mapping \citep{peterson1993}. As radii derived from reverberation mapping are available only for a rather small number of AGN, the monochromatic luminosity of the AGN continuum is commonly used as a proxy for the radius. The virial mass estimator then takes the form
\begin{equation}
M = \kappa\,v^2\,(\lLl)^{\alpha}
\label{eq:mass}
\end{equation}
where $\Ll$ is the specific continuum luminosity at wavelength $\lambda$, $\kappa$ is a constant to be calibrated, and $\alpha\approx0.5$ \citep{kaspi2007, bentz2009}. The monochromatic luminosities $\lLl$ are estimated from the continuum close to the emission line from which $v$ is derived. Estimators of this type, for various lines, have been used successfully for deriving black hole masses for large samples of AGN \citep{greene2005, vestergaard2006, kim2010, park2013, jun2015}.

Currently used mass estimators are based on the expectation that the best proxy for the radius of the broad line region for a given line is $R\propto(\lLl)^{\alpha}$. A priori however, this is an ad-hoc assumption that requires justification, and alternative scalings might be possible. In the following, I thus explore the hypothesis that the sizes of broad line regions do not scale with $\lLl$ but rather with \emph{bolometric} luminosity.

\section{Color Effects \label{sec:color}}

I begin with the assumption that the correct mass estimator based on an optical/ultraviolet emission line $i$ is
\begin{equation}
M_i = \kappa_i\,v^2_i\,L^{\alpha}
\label{eq:mass0}
\end{equation}
where $L$ is the \emph{bolometric} luminosity of the optical/ultraviolet AGN continuum and $\alpha\approx0.5$ for all lines. At least in some cases, we can analyze the same target using two lines $a, b$, resulting in two mass estimates $M_a, M_b$. Provided that both mass estimators have been calibrated correctly, the results for $M_a$ and $M_b$ should be identical (within errors) for any given target, meaning
\begin{equation}
\label{eq:mass0ratio}
\frac{M_a}{M_b} = \frac{\kappa_a}{\kappa_b} \left(\frac{v_a}{v_b}\right)^2 = 1 \ .
\end{equation}

If we replace the bolometric luminosity $L$ by the monochromatic luminosity derived from a narrow band centered on a wavelength $\lambda$ located close to the emission line to be analyzed, $\lLl$, we find a modified mass estimator
\begin{equation}
M'_i = \kappa_i\,v^2_i\,\xi_i\,L_i^{\alpha}
\label{eq:mass1}
\end{equation}
with $L_i\equiv(\lLl)_i$ for convenience and $\xi_i$ being another constant. When analyzing the same target using two different lines $a, b$, we find
\begin{equation}
\label{eq:mass1ratio}
\frac{M'_a}{M'_b} = \frac{\kappa_a}{\kappa_b} \left(\frac{v_a}{v_b}\right)^2 \frac{\xi_a}{\xi_b} \left(\frac{L_a}{L_b}\right)^{\alpha} .
\end{equation}
Applying Equation~(\ref{eq:mass0ratio}), we can reduce Equation~(\ref{eq:mass1ratio}) to
\begin{equation}
\label{eq:mass1ratio2}
\frac{M'_a}{M'_b} = \frac{\xi_a}{\xi_b} \left(\frac{L_a}{L_b}\right)^{\alpha} = \xi_{ab} \left(\frac{L_a}{L_b}\right)^{\alpha}
\end{equation}
using here $\xi_{ab}\equiv(\xi_a/\xi_b)$ for convenience.

Assuming that (i) both mass estimators $M_{a,b}$ have been calibrated independently and using sufficiently representative AGN samples, and (ii) the ratio $M'_a/M'_b$ can be derived for a sufficiently large and representative sample of AGN, the \emph{ensemble average} of the mass ratio, $\langle M'_a/M'_b\rangle$, will be unity (within errors) by construction:
\begin{equation}
\label{eq:mass1meanratio}
\left\langle\frac{M'_a}{M'_b}\right\rangle = \left\langle \xi_{ab} \left(\frac{L_a}{L_b}\right)^{\alpha} \right\rangle = \xi_{ab} \left\langle \left(\frac{L_a}{L_b}\right)^{\alpha} \right\rangle = 1 \ .
\end{equation}
This relation implies $\xi_{ab}=\langle(L_a/L_b)^{\alpha}\rangle^{-1}$, at least if the values $M'_a/M'_b$ are derived for a sufficiently large and representative sample. In case of small samples (as frequently the case), $\xi_{ab}$ needs to be estimated separately from the typical optical/ultraviolet color of the AGN continuum.

\section{Comparison to Observations \label{sec:obs}}

Arguably the most careful cross-analysis of different mass estimators is provided by \citet{assef2011} who compare mass estimates based on the \ion{C}{iv}\,$\lambda$1549 line to those based on the hydrogen Balmer lines \Ha\,$\lambda$6565 and \Hb\,$\lambda$4863 (with all wavelengths in \AA). \citet{assef2011} apply the mass estimators
\begin{eqnarray}
\label{eq:assefmasses}
\frac{M'\Hbb}{\Msol} \hskip-5pt &=& \hskip-5pt 6.71\hskip-3pt\times\hskip-3pt 10^6 f \left(\frac{v\Hbb}{10^6\,{\rm m\,s^{-1}}}\right)^2 \left(\frac{\LH}{10^{37}\,{\rm W}}\right)^{0.52} \nonumber \\
\frac{M'\Haa}{\Msol} \hskip-5pt &=& \hskip-5pt 7.68\hskip-2pt\times\hskip-2pt 10^6 f \left(\frac{v\Haa}{10^6\,{\rm m\,s^{-1}}}\right)^{2.06} \hskip-2pt \left(\frac{\LH}{10^{37}\,{\rm W}}\right)^{0.52}  \nonumber \\
\frac{M'\C}{\Msol} \hskip-5pt &=& \hskip-5pt 10^{\epsilon} \, \left(\frac{v\C}{10^6\,{\rm m\,s^{-1}}}\right)^2 \left(\frac{\LC}{10^{37}\,{\rm W}}\right)^{0.53} ,
\end{eqnarray}
drawn from \citet{greene2005}, \citet{vestergaard2006}, and \citet{bentz2009}, to a sample of 12 lensed high-redshift quasars. Here $\LC$ and $\LH$ denote the continuum luminosities $\lLl$ derived at 1350\,\AA\ and 5100\,\AA, respectively; the values of the constants $f$ and $\epsilon$ depend on if $v$ is derived from the dispersion or the FWHM of a line. The relation between $M'\Haa$ and $v\Haa$ is actually $M'\Haa\propto v\Haa^{2.06\pm0.06}$, i.e., consistent with $M\propto v^2$. Likewise, the power law indices of $\LC$ and $\LH$ are consistent with 0.5 within errors.

From their analysis, \citet{assef2011} find a systematic discrepancy between the masses derived from the Balmer lines on the one hand and from \ion{C}{iv} on the other hand. This discrepancy follows the relation (their Equation~(8))
\begin{equation}
\frac{M'\H}{M'\C} = 10^{-y} \left(\frac{\LC}{\LH}\right)^{-x}
\label{eq:assefratio}
\end{equation}
where the subscript H denotes either \Ha\ or \Hb. Depending on which Balmer line and which velocity indicator (line dispersion or line FWHM) is used, best-fit values range from $0.51\pm0.14$ to $0.95\pm0.22$ for $x$ and $-0.11\pm0.06$ to $-0.27\pm0.08$ for $y$.

If indeed the broad line region radius scales with bolometric luminosity rather than with $\lLl$, i.e., if Equation~(\ref{eq:mass0}) gives the true underlying relation, then Equations (\ref{eq:mass1ratio2}) and (\ref{eq:assefratio}) are equivalent. This can be checked in a straightforward manner by identifying either \Ha\ or \Hb\ with ``line $a$'' and \ion{C}{iv} with ``line $b$''. Evidently, Equation~(\ref{eq:mass1ratio2}) predicts $\alpha=x$. Indeed, the values observed by \citet{assef2011} for $x$ are in agreement with $\alpha\approx0.5$ within errors.

The \emph{ensemble averaged} optical/ultraviolet continuum of quasars is known to follow the relation $\lLl\propto\lambda^{-0.56}$ in the wavelength range 1300--5500\,\AA\ \citep{vandenberk2001}. Using $\alpha=0.5$, this   implies\footnote{I use here $\langle(L_a/L_b)^{\alpha}\rangle^{-1} \approx \langle L_a/L_b\rangle^{-\alpha}$. This is possible because the ratio $L_a/L_b$ is of order unity always and $0.56\alpha<1$.}   $\xi_{ab} \approx [(5100\,{\textrm\AA}/1350\,{\textrm\AA})^{-0.56}]^{-0.5} \approx 1.45$. Comparison of Equations (\ref{eq:mass1ratio2}) and (\ref{eq:assefratio}) leads to the prediction $\log\xi_{ab}\approx0.16=-y$. Within errors, this is in agreement with the values observed by \citet{assef2011}.

\section{Discussion and Conclusions \label{sec:discuss}}

From the analysis provided in Section~\ref{sec:obs}, it is straightforward to see that the observed systematic discrepancy between mass estimates based on different emission lines is in agreement with AGN broad line regions scaling with bolometric rather than monochromatic luminosity. On the one hand, the fact that AGN broad line regions are shaped by both the optical/ultraviolet and the ionizing Lyman continuum was already noted by \citet{peterson1993} and \citet{baldwin1995}. On the other hand, little effort has been made so far to distinguish scalings with $L$ from those with $\lLl$ observationally. It now seems that the data of \citet{assef2011} provide the as yet most direct evidence for a scaling of the radii of AGN broad line regions with bolometric luminosity.

Comparisons of black hole masses derived from optical and ultraviolet lines usually focus on the line widths rather than the broad line region radii (cf., e.g., \citealt{ho2012} vs. \citealt{runnoe2013}); this might explain why color effects have largely been overlooked so far. Even though, an independent hint might have been provided by the \emph{Baldwin effect} \citep{baldwin1977}, a characteristic anticorrelation between the equivalent widths of ultraviolet emission lines and continuum luminosity: whereas equivalent widths are usually discussed as function of $\lLl$, the actual underlying correlation might be the one between equivalent width and bolometric luminosity \citep[cf.,][and references therein]{xu2008}.

When expressed as functions of bolometric AGN luminosity (to ease comparisons of different lines), characteristic broad line region radii can be assigned to individual emission lines; for the case of \Hb\ vs. \ion{C}{iv}, the radii of the former are about three times larger than the radii of the latter in general (cf. Chapter~7.1.8 of \citealt{netzer2013}). Such a gas stratification is the natural consequence of the differences in ionization potential; lines requiring higher excitation or ionization energies have to be located closer to the central source of radiation. Indeed, studies of the radius--luminosity relationships of broad line regions assume (at least implicitly) that $\lLl$ is a sufficient proxy for the bolometric continuum luminosity \citep[e.g.,][]{kaspi2007, bentz2009}. Such studies derive radius--luminosity relationships, as well as black hole mass--luminosity relationships, by analyzing radius or mass as function of $\lLl$ and applying global scaling factors that implicitly include an \emph{ensemble averaged} bolometric correction. However, if broad line regions scale with bolometric AGN luminosity, bolometric corrections need to be applied \emph{source by source}. As demonstrated by \citet{assef2011}, neglecting colors and bolometric corrections leads to differences up to a factor of four between black hole masses estimated from the Balmer lines and \ion{C}{iv}, respectively. This also suggests that \emph{all} emission-line based mass estimators are biased to varying degrees.

Assuming a scaling of mass estimates with bolometric luminosity, the cross-comparison of non-simultaneous virial mass estimates based on different lines is affected by the variability of the AGN luminosity. Quasars are known to vary in luminosity by factors up to about three on time scales of years \citep[e.g.,][]{schramm1993}; AGN variability in general obeys \emph{red noise} statistics, meaning that stronger variations occur on longer time scales \citep{park2012, park2014, kim2013}. However, such variability cannot introduce systematic trends into samples of sources; rather, it increases the measurement error for individual sources. In general, variability effects are relatively moderate because only the square root of the luminosity enters the mass estimate. The individual mass estimates used by \citet{assef2011} come with errors of around 0.3~dex; accordingly, variations in luminosity up to factors of about 5 would be within the measurement errors (because $\log(\sqrt{5})=0.35$).

The discussion provided in this paper, along with the one by \citet{assef2011}, illustrates once more the fact that any systematic scatter in a given relation, like the one between mass estimates based on \Ha/\Hb\ and \ion{C}{iv}, is indicative of hidden parameters. Only a careful analysis of the impact of AGN continuum color eventually leads to the insight that AGN broad line regions scale with bolometric luminosity rather than $\lLl$. (Another recent example is provided by the relation between Bondi accretion rate and kinetic jet power in radio galaxies; \citealt{trippe2014}.) In turn, blind application of ``standard'' relations despite the presence of obvious systematic scatter comes with the risk of producing results that are seriously biased.


\acknowledgments

I am grateful to {\sc\small Marios Karouzos} (SNU Seoul) for pointing out the work by \citet{assef2011} to me and for valuable discussion, and to an anonymous referee for an encouraging report. I acknowledge financial support from the Korean National Research Foundation (NRF) via Basic Research Grant 2012-R1A1A-2041387.



\end{document}